\documentclass{article}
\usepackage[dvips]{graphicx}

\begin{document}

\vspace*{1.0cm}
\centerline{\LARGE \bf A solvable  quantum antiferromagnet model}
\vspace*{0.25cm}

\begin{center}
{\large Bikas K. Chakrabarti\,\ddag\dag \,\,\,and Jun-ichi Inoue\,\dag} 
\mbox{}\\
\ddag \, Saha Institute of Nuclear Physics, 1/AF Bidhannagar, 
Kolkata 700064, India. \\
{\tt bikask.chakrabarti@saha.ac.in}\\
\dag \, Graduate School of Information 
Science and Technology, Hokkaido University,\\ 
N14-W9, Kita-ku, Sapporo 060-0814, Japan.\\
{\tt j\_inoue@complex.eng.hokudai.ac.jp}

\end{center}

\begin{abstract}
We introduce a  quantum 
antiferromagnet model, having exactly soluble thermodynamic
properties. It is an infinite range antiferromagnetic Ising model
put in  a transverse field.  The free energy gives the ground state
energy in the zero temperature limit and it also gives the low
temperature behaviour of the specific heat, the exponential variation
of which gives the precise gap magnitude in the excitation spectrum of
the system. The detailed behaviour of the (random sublattice) staggered
magnetisation and susceptibilities are obtained and studied near the N\'eel
temperature and the zero temperature quantum critical point.
\end{abstract}

\vskip 1 cm
\noindent
With the realisation in the mid last century, that the N\'eel state can not
be the ground state, or even an eigenstate, of the quantum Heisenberg 
antiferromagnets, considerable efforts have gone in search for the 
nature of the ground state for such quantum antiferromagnets. Searches
have also been made for low lying eigen states of Heisenberg antiferromagnets
 even in the semi-classical limits, or in some other variants of the model
where exact ground state 
and other eigenstates are available [1,2]. In particular,
the exact dimerised (two-fold degenerate; disordered) ground state in the
one dimensional next nearest neighbour interacting Heisenberg antiferromagnet
(having strength ratio 1/2 between
the  next nearest and nearest neighbours) has been
found out [3] and this  observation has attracted considerable 
attention in the context of high temperature superconductors occurring
in materials having antiferromagnetic properties [1,2]. Regarding the
excited states in such quantum antiferromagnets, the Haldane conjecture
[2,4] states that the integer spin systems are massive (have a gap), while
the half-integer ones are massless.   

The low temperature limit of the exact free energy 
obtained here for  the (spin-${1\over 2}$) long-range interacting
transverse Ising model shows that the specific heat has an exponential 
(in gap and inverse temperature) variation, giving the precise 
 magnitude of the gap in the model. 
 The order-disorder transition, driven both by temperature and the tunneling
or transverse field, are investigated studying  the ordering and 
susceptibility behaviours.

The model we study here has the Hamiltonian
\begin{eqnarray}
 H = {\frac{J}{N}}
\sum _{i,j} S^z_i S^z_j - h \sum _i S_i^z -
\Gamma \sum _i S_i^x, 
\end {eqnarray}
where $J$ denotes the long-range antiferromagnetic ($J > 0$) exchange
constant and $S^x$ and $S^z$  denote the $x$ and $z$ components
of the $N$ Pauli spins ($S$ = 1/2):

$$ S^x_i = \left(
\begin{array}{cc}
0 & 1 \\
1 & 0
\end{array}
\right), \,\,\,
S_{i}^z =
\left(
\begin{array}{cc}
1 & 0 \\
0 & -1
\end{array}
\right), \,\,\,
 i = 1, 2, ..., N.$$
Denoting half of the randomly chosen lattice sites as members of 
sublattice $A$ and the rest of $B$, and expressing the cooperative
term in Hamiltonian (1) as the difference of two quadratics, the Hamiltonian
can be rewritten as
\begin{eqnarray}
 H & = & {\frac{J}{2N}} \left (
\left[ \sum _{i} S^z_{i(A)} + \sum _{i}  S^z_{i(B)} \right]^2
- \left[ \sum _{i} S^z_{i(A)} - \sum _{i}  S^z_{i(B)} \right]^2\right) 
 \nonumber \\
& &
- h \sum _i (S_{i(A)}^z +   S^z_{i(B)}) -
\Gamma \sum _i (S_{i(A)}^x +   S^x_{i(B)}). \nonumber
\end{eqnarray}
In fact, the Hamiltonian (1) is meaningful when the spins in the cooperative
term there belongs to two different (though randomly defined) sublattices,
each consisting of $N/2$ spins.
Using the Hubbard-Stratonovich transformation, the partition function can 
be expressed as
\begin{eqnarray}
Z\mbox{} & = &
{\rm Tr}_{
\mbox{ $S$}_{A},
\mbox{ $S$}_{B}}
\int_{-\infty}^{\infty} \int_{-i\infty}^{+i\infty}
{\frac{idm_{+}dm_{-}} {(2\pi/N\beta J)}}
{\rm exp}\left [{\frac{N\beta J}{2}}(m_{+}^{2} -m_{-}^{2})\right] \nonumber \\
\mbox{} & \times & {\exp} {\left [ (\beta J m_{+}) \sum_{i}
(S_{i(A)}^{z}+ S_{i(B)}^{z}) + (\beta J m_{-}) \sum_{i}
( S_{i(A)}^z -    S^z_{i(B)}) \right .}\nonumber \\
\mbox{} & + & {\left .\beta h \sum _i (S_{i(A)}^z +  S^z_{i(B)})
+ \beta \Gamma
\sum _i (S_{i(A)}^x +   S^x_{i(B)}) \right ]},
\end{eqnarray}
where $\beta $ denotes the inverse temperature $T$ and $m_+$,
 $m_-$ are related to the (uniform; $z$-component) magnetisation
$M^z_A$ and $M^z_B$ of the sublattices $A$ and $B$ respectively: $m_+ =
- (M^z_A + M^z_B)$ and $m_- = (M_A^z - M_B^z)$. The trace in (2) can now be
easily performed [5] as each spin {$\vec S$} can now be imagined to be isolated
and present only in a vector field {$\vec h$} having both $\-z$ and $\-x$
components: $h^x = \Gamma$ and $h^z \sim \beta J m$. In the $N \rightarrow
\infty$ limit, the free energy $f$ per spin can be obtained from 
\begin{eqnarray}
 Z & = & {\exp} [- N \beta f], \nonumber \\
 f & = & {\frac {J}{2}} \left [ (M_A^z + M_B^z)^2 -
 (M_A^z -M^z_B)^2 \right] \nonumber \\
\mbox{}   & + & {\frac {1}{2 \beta}} {\log} \left[
\cosh \beta \sqrt{ (2JM_{B}^{z}-h)^{2} + \Gamma^{2}} \right]   
 + {\frac {1}{2 \beta}}
\log \left[ \cosh \beta \sqrt{ (2JM_{A}^{z}-h)^{2}+ \Gamma^{2}} \right],
\end{eqnarray}
where the sublattice magnetisations $M^z_A$ and $M^z_B$ are given by
the (self-consistent) saddle point equations:
\begin{eqnarray}
M_{A}^{z} & = & \frac{(-2JM_{B}^{z}+h)}
{\sqrt{(-2JM_{B}^{z}+h)^{2}+\Gamma^{2}}}
\tanh \beta
{\sqrt{(-2JM_{B}^{z}+h)^{2}+\Gamma^{2}}} \nonumber\\
M_{B}^{z} & = &
\frac{(-2JM_{A}^{z}+h)}
{\sqrt{(-2JM_{A}^{z}+h)^{2}+\Gamma^{2}}}
\tanh \beta
{\sqrt{(-2JM_{A}^{z}+h)^{2}+\Gamma^{2}}}.
\end{eqnarray}

\begin{figure}
\begin{center}
\resizebox{6.5cm}{!}{\includegraphics{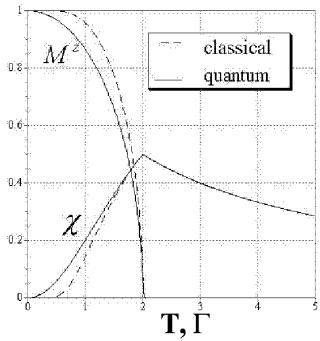}}
\caption{\small{Variations of longitudinal sublattice magnetization
and susceptibilities in the 
classical and quantum cases ($J = 1$).}}
\label{Fig-mz}
\end{center}
\end{figure}

The spontaneous sublattice order $M^z_A$ or $M^z_B$ vanishes
at the N\'eel phase boundary $T_N (\Gamma)$; see Fig.~\ref{Fig-mz}. It may be 
noted that the choice of sublattices
here is purely random and the sublattice order 
therefore looks like that of a  glass. Also,
deep inside the antiferromagnetic phase (at 
$\beta \rightarrow \infty, \Gamma \rightarrow 0, h = 0$), $M_A^z =
1 = -M^z_B$, so that the free energy $f$ can be expressed as $f \sim ({1}/
{\beta}) \log [1 + \exp (-2\beta \Delta)]$ and the specific heat $\partial^2 f/
\partial T^2 $ will have a variation like $\exp [-2\beta \Delta (\Gamma)]$,
similar to those of a two level
system with a gap $\Delta (\Gamma) = {\sqrt {4J^2 + \Gamma^2}}$. 
This is the
exact  magnitude of the gap in the magnon spectrum of this long range
transverse Ising antiferromagnet. Actually, the above result for the gap
can be seen directly from the effective Hamiltonian in (2) having a simple
$\left(
\begin{array}{cc}
a & b \\
b & -a
\end{array}
\right)$
matrix form where $a = 2\beta J m$
and $b = \Gamma$, giving the eigen values $\pm \beta\sqrt{4J^2m^2 + \Gamma^2}$.
This gives the gap $\beta \Delta$ at $T= 0 =\Gamma$, where the  
magnetisation $m $ becomes equal to unity.

In the classical case, near the  N\'eel temperature $T_N^0 \equiv 
T_N(\Gamma = 0)$, 
we expand the equations (4)   
around 
$M_{A} \simeq 0, M_{B} \simeq 0$ for 
$h=0$ and obtain 
$M_{A}^{z} \simeq -2J\beta M_{B}^{z} +\mathcal{O} ((M_B^z)^3), 
M_{B}^{z} \simeq -2J\beta M_{A}^{z}  +\mathcal{O} ((M_A^z)^3)$. 
The only possible solutions for these  linearized equations are  
$M_{A}^{z}= -M_{B}^{z}=M^z = 0$ and
$M_{A}^{z}=-M_{B}^{z}= M^z \sim (T - T_N^0)^{1/2}$, where $T^0_{N}=2J$. 
The longitudinal linear susceptibilities $\chi_{A/B}$  
are given by $\chi_{A/B}  =  
\lim_{h \to 0} {\partial M_{A/B}^{z}}/{\partial h} =  
{\beta (1- 2J\chi_{B/A})} {\cosh^{2} \beta (-2J M_{B/A}^{z})}$.
Hence, $\chi = \chi_A +   \chi_B$ behaves as 
\begin{eqnarray}
\chi & = & \frac{2}{T^0_N + T \cosh^{2} (2JM^z)/T}.
\end{eqnarray}
This gives 
$\chi  =   {2}/({T+T^0_{N}})$
for high temperatures $T > T_N^0$ (where $M_A =- M_B= M^z =0 $) and at $T_N^0$
it grows upto a value $1/2J$ and drops down again to 0 as $T \to$ 0, giving
the well known cusp behaviour (see e.g., [6]) for the classical 
antiferromagnets. 

\begin{figure*}
\begin{center}
\resizebox{6.5cm}{!}{\includegraphics{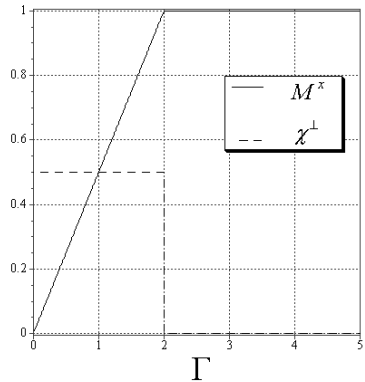}}
\caption{\small{Variations of transverse magnetisation and susceptibility
 in the quantum case ($J = 1$).}}
\label{Fig-chiperp}
\end{center}
\end{figure*}

In the pure quantum case ($T = 0$, $\Gamma \ne 0$), the $\tanh $ term in 
(4) equals to unity and with a similar expansion near the quantum critical
point $\Gamma_N = 2J$, one gets the longitudinal susceptibility $\chi
= \chi_A + \chi_B$; $\chi_{A/B} = 
\lim_{h \to 0}
{\partial M_{A/B}^{z}}
/{\partial h} = 
{\Gamma^{2}
(1-2J\chi_{B/A})}
/{[(2JM_{B/A}^{z})^{2}+\Gamma^{2}]^{3/2}}.$
We then have  
\begin{eqnarray}
\chi & = & \frac{2\Gamma^{2}} {[(\Gamma_N M^z)^{2}+
\Gamma^{2}]^{3/2} + \Gamma_N \Gamma^{2}},
\end{eqnarray}
giving 
$\chi  = {2}/({\Gamma + \Gamma_{N}}) $ for $\Gamma > \Gamma_N$  and growing
upto a value $1/2J$ at $\Gamma = \Gamma_N$ and then decreasing eventually to
a value $\chi =  \Gamma^2/4J^3$ as $\Gamma \to 0$. Again a cusp 
behaviour is seen for the susceptibility at the quantum critical point 
$\Gamma_N$ for such a quantum antiferromagnet (see Fig.~\ref{Fig-mz}). 
The behaviour is of course qualitatively similar 
to that the classical critical (N\'eel) point.
One can also study the transverse susceptibility $\chi^{\perp} (=
dM^x/d\Gamma)$ behaviour. One finds
$\chi^{\perp}_A =0=  \chi^{\perp}_B$ for $\Gamma > \Gamma_N$ and 
$\chi^{\perp}_A = 
1/2J = \chi^{\perp}_B$ for the ordered phase (see Fig.~\ref{Fig-chiperp}).

In view of the intriguing ground state properties of quantum
 antiferromagnets [2], indications coming often
 only from approximate theories (see e.g., [7] for
a long range quantum Heisenberg antiferromagnet) or numerical
simulations (see e.g., [8] and references therein), our
proposed quantum
antiferromagnetic model, having exactly soluble thermodynamic
properties, should be of some interest.
 The model consists of an  infinite range antiferromagnetic 
Ising system, 
put in  a transverse field.  The classical ground state of the model is 
highly degenerate. Although no signature of slow dynamics, like in glasses,
can be seen here, the ordered state in the system corresponds to 
(quantum) glass-like system as well. The number of degenerate states
can be estimated to be $\mathcal {O} (2^{N/2})$, which is larger than that for 
the Sherrington-Kirkpatrick model ($\sim \mathcal {O}(2^{0.28743N}))$ [9].  
 This may be compared and contrasted
with  the transverse Ising antiferromagnets on topologically
frustrated triangular lattices studied extensively in the last
few years [10].
 The free energy in (3) gives the ground state
energy in the zero temperature limit and it also gives the low
temperature behaviour of the specific heat, the exponential variation
of which gives the precise gap magnitude 
$\Delta (= {\sqrt{ 4J^2 + \Gamma^2}}) $ in the excitation spectrum of
the system. It may be noted that although it is a
spin-1/2 system, because of the
restricted (Ising) symmetry and the infinite dimensionality (long
range interaction) involved, there need not be any  conflict
with the Haldane conjecture. Although our entire analysis has been for
spin-1/2 (Ising) case, because of the reduction of the effective
Hamiltonian in (2) to that of a single spin in an effective vector field, the
results can be easily generalised for higher values of the spin $S$. No
qualitative change is observed. 
 The order-disorder transition in the model can be driven both by
thermal fluctuations (increasing $T$) or by the quantum fluctuations
(increasing $\Gamma$). These transitions in the model have been investigated
here 
studying the  behaviours of the (random sublattice) 
magnetisation and the (longitudinal and transverse) susceptibilities.
No quantum phase transition, where the gap $\Delta$ vanishes, is observed
in the model, unlike in the one dimensional transverse Ising antiferromagnets
[1,5].

\vskip 1 cm
\noindent {\bf Acknowledgements:}
The  work was financially 
supported by 
{\it IKETANI Science \& Technology Foundation} grant no.
0174004-D. 
One of the authors (B. K. C.) 
thanks University of Tokyo and Hokkaido University
for their hospitality. We are also grateful to S. M. Bhattacharjee,
I. Bose and Y. Sudhakar for illuminating discussions.


\vspace{5.0mm}


\begin{thebibliography}{99}

\bibitem {Sachdev} 
See e.g., S. Sachdev, {\it Quantum Phase Transitions}, Cambridge
Univ. Press, Cambridge (1999). 

\bibitem {Fradkin} See e.g.,  E. Fradkin, {\it Field Theories of Condensed Matter
Systems}, Addison-Wesley, Redwood (1991); I. Bose, in {\it Frontiers in 
Condensed Matter Physics: 75th Year Special Publication of Indian
Journal of Physics}, Vol. 5, Eds. J. K. Bhattacharjee and B. K. Chakrabarti,
Allied Publishers, New Delhi (2005).

\bibitem {CKM} C. K. Majumdar and D. K. Ghosh, J. Math. Phys. {\bf 10}
1388, 1399 (1969).

\bibitem {Haldane} F. D. M. Haldane, Phys. Lett. A {\bf 93} 464 (1983);
Phys. Rev. Lett. {\bf 50} 1153 (1983); I. Affleck, T. Kennedy, E. H. Lieb
and H. Tasaki, Phys. Rev. Lett. {\bf 59} 799 (1987).


\bibitem{Bikas}
B. K. Chakrabarti, A. Dutta and 
P. Sen, 
{\it Quantum Ising Phases and 
Transitions in Transverse Ising Models}, 
Springer, Heidelberg (1996). 

\bibitem{Kittel}
C. Kittel, 
{\it Introduction to 
Solid State Physics}, 
John Wiley $\&$ Sons Inc., N. Y. (1966).

\bibitem{Kaiser}
C. Kaiser and I. Peschel, 
J. Phys. A : 
Math. Gen. {\bf 22} 4257 (1989). 

\bibitem{Roscilde}
T. Roscilde, P. Verrucchi, A. Fubini, S. Haas and 
V. Tognetti, 
Phys. Rev. Lett. {\bf 94}, 147208 (2005). 

\bibitem{Tanaka}
F. Tanaka and S. F. Edwards, J. Phys. F: Metal Phys. {\bf 10}, 2789 (1980).

\bibitem{Moessner}
R. Moessner, S. L. Sondhi and P. Chandra, Phys. Rev. Lett. {\bf 84} 
4457 (2000); 
R. Moessner and S. L. Sondhi, 
Phys. Rev. B {\bf 63}, 224401 (2001). 

\end{thebibliography}
\end{document}